# Glassy Dynamics Under Superhigh Pressure

A.A. Pronin[1], M.V. Kondrin[2], A.G. Lyapin[2], V.V. Brazhkin[2], A.A. Volkov[1], P. Lunkenheimer[3,*], and A. Loidl[3]

[1] *General Physics Institute, Russian Academy of Sciences, Moscow, Vavilov st. 38, 119991, Russia*
[2] *Institute for High Pressure Physics, Russian Academy of Sciences, Troitsk, Moscow region, 142190, Russia*
[3] *Experimental Physics V, University of Augsburg, 86159 Augsburg, Germany*

Nearly all glass-forming liquids feature, along with the structural $\alpha$-relaxation process, a faster secondary process ($\beta$-relaxation), whose nature belongs to the great mysteries of glass physics. However, for some of these liquids, no well-pronounced secondary relaxation is observed. A prominent example is the archetypical glass-forming liquid glycerol. In the present work, by performing dielectric spectroscopy under superhigh pressures up to 6 GPa, we show that in glycerol a significant secondary relaxation peak appears in the dielectric loss at $P > 3$ GPa. We identify this $\beta$-relaxation to be of Johari-Goldstein type and discuss its relation to the excess wing. We provide evidence for a smooth but significant increase of glass-transition temperature and fragility on increasing pressure.

Under sufficiently fast cooling or compression, many liquids do not crystallize but transfer into a structurally disordered glassy state. Various attempts at theoretically describing this phenomenon have been made during the last century [1] but this problem still is a major scientific challenge. The permittivity spectra of supercooled liquids reveal several relaxation features, which are believed to be the key to achieve a better understanding of the glass transition. The most prominent of them is the structural $\alpha$-relaxation characterized by an exponentially strong dependence of the relaxation time on temperature and pressure, traced over 15-18 decades of frequency [2,3]. In addition, all glass formers reveal faster secondary (or $\beta$-) relaxation processes. They appear as a well-defined peak at higher frequencies than the $\alpha$-peak in dielectric loss spectra (type B systems [4]) or as an excess wing (EW) on the high-frequency flank of the structural relaxation (type A). Secondary relaxations termed Johari-Goldstein (JG) relaxation [5] are believed to be intrinsic properties of supercooled liquids but their microscopic origin is still controversially discussed. JG relaxations should be distinguished from relaxations caused by intramolecular motions, which are of minor interest for the understanding of the glass transition. A detailed classification of secondary relaxation processes is found in [6]. Also the origin of the EW is debatable: For example, in [7,8,9,10] it was suggested that it is the high-frequency flank of the JG-type secondary relaxation peak hidden under the $\alpha$-peak. On the other hand, the EW relaxation may also be a separate phenomenon, not related to the JG relaxation [4].

The application of high pressure is very advantageous for investigating glassy dynamics because different relaxation modes have different sensitivity to pressure [11,12]. In the case of the EW, using pressure as a controlling parameter is of special importance: the EW synchronously shifts with the $\alpha$-peak when temperature changes so that they cannot be distinguished on the basis of temperature alone.

Glycerol is the most extensively studied glass-forming liquid. Broadband dielectric loss spectra [3] reveal it to be a typical type A system. In several type-A glass formers, the EW was found to develop into a weak shoulder after several weeks aging [9,13] indicating that it is due to a secondary relaxation. However, in glycerol there are only very faint indications for such a shoulder [9]. In addition, in contrast to other glass formers [12,14], under elevated pressures the secondary relaxation in glycerol does not become sufficiently strong to show up as a clear shoulder or peak [10,15,16,17]. First measurements on glycerol under pressures up to 5 GPa were performed 40 years ago [17]. However, the employed high-pressure technology did not allow obtaining reliable high-quality spectra at pressures above 2 GPa. Since then, several attempts have been made to study glycerol under pressure [10,15,16,18,19], however, with pressures up to 1-2 GPa only. Interestingly, recent numerical results obtained within mode coupling theory [20] predict a crossover from a "temperature-driven" to a "density-driven" glass transition at very high pressures, which in glycerol may occur for pressures beyond 3 GPa.

For the dielectric measurements a toroid-type high pressure cell was developed [21]. Due to the highly uniform deformation of a liquid-filled container, it allows attaining quasi-hydrostatic pressures with very little shear strain even after the solidification of the liquid under study. Temperature and pressure were measured by a thermocouple and manganin manometer, both placed in the vicinity of the sample. For impedance measurements at frequencies 10 Hz - 2 MHz, a QuadTech 7600 precision LCR meter was used. In the employed setup, the measured absolute values of the real part of the permittivity have some uncertainties due to possible small contributions from stray capacitance. Samples of ultra-pure (>99%) glycerol were purchased from ICN Biomedicals. Sample purity level as determined by refractometry was better than 99%. In addition, the sample spectra were checked in situ at ambient conditions each time prior to the pressure experiments to exclude possible contamination at a cell assembly stage.

Since in the pressure range above several GPa the intermolecular repulsion energy becomes comparable with the binding energy inside molecules, it is appropriate to consider possible chemical instability of glycerol under such extreme conditions. To exclude this possibility, the glycerol sample was subjected to a pressure as high as 6 GPa at 310 K; then the pressure was released and temperature-dependent dielectric spectra were registered. The results remained unchanged within experimental error, showing no signs of irreversible chemical transformations such as polymerization or decomposition.

$$\varepsilon^* = \varepsilon' - i\varepsilon'' = \varepsilon_\infty + \Delta\varepsilon_{CD}/(1+i\omega\tau_{CD})^{\beta_{CD}}. \quad (1)$$

We used the Cole-Cole (CC) expression [23],

$$\varepsilon^* = \varepsilon_\infty + \Delta\varepsilon_{CC}/\left[1+(i\omega\tau_{CC})^{1-\alpha_{CC}}\right] \quad (2),$$

for modeling the $\beta$-peak, and the sum of (1) and (2) when both features are simultaneously observed.

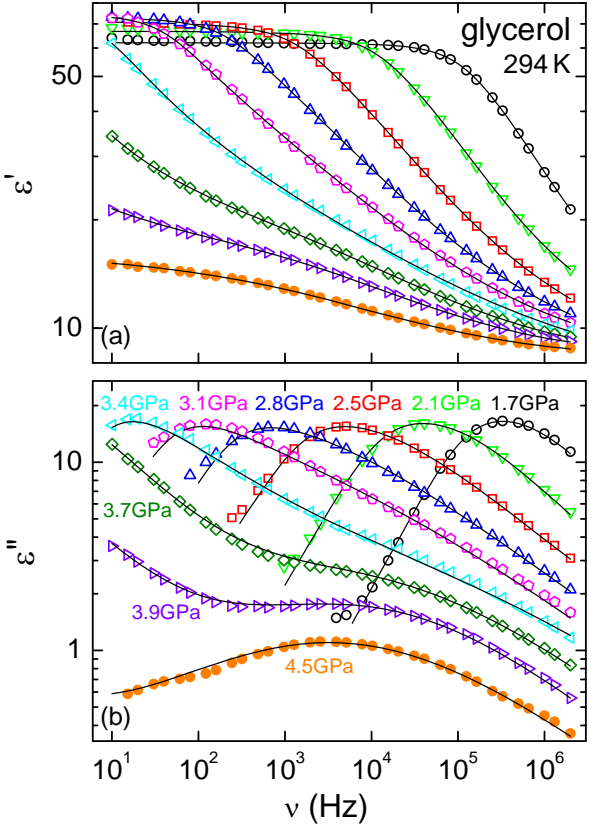

FIG. 1 (color online). Glycerol spectra at different pressures and 294 K. The lines are fits with the sum of the CD model, eq. (1), for the $\alpha$-peak and the CC model, eq. (2), for the $\beta$-peak.

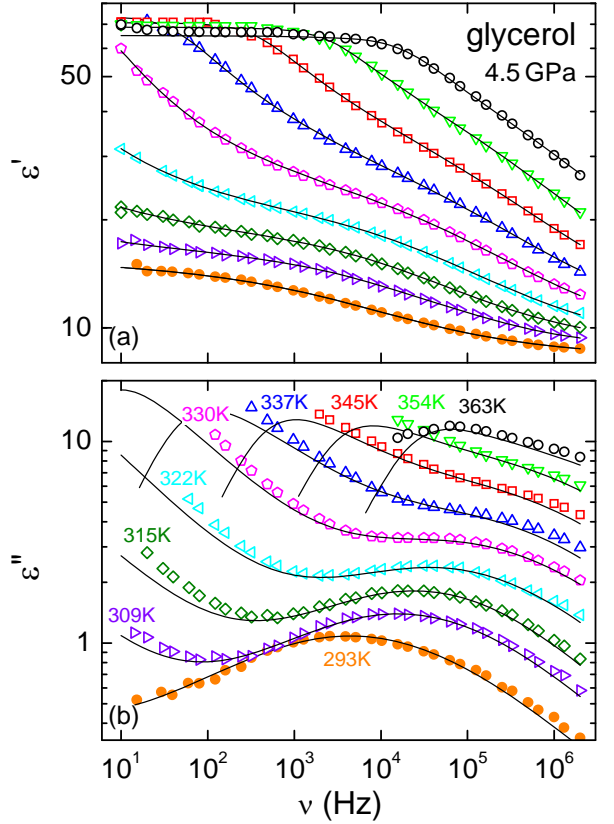

FIG. 2 (color online). Glycerol spectra under 4.5 GPa at various temperatures. The lines are fits with the sum of the CD model, eq. (1), for the $\alpha$-peak and the CC model, eq. (2), for the $\beta$-peak. The significant conductivity contributions showing up in $\varepsilon''(\nu)$ at low frequencies have been neglected in the fits.

Figures 1 and 2 show representative results from our high-pressure investigations, namely isothermal spectra of the real and imaginary part of the permittivity for different pressures at room temperature (Fig. 1) and isobaric spectra for different temperatures at 4.5 GPa (Fig. 2). The most remarkable result is the emergence of a distinct secondary relaxation feature, starting from pressures of about 3 GPa. This finding unequivocally shows that also in the prototypical type-A glass former glycerol a secondary relaxation is present. The $\alpha$-relaxation can be well described by the empirical Cole-Davidson (CD) expression [22]:

Figure 3 shows the pressure dependences of the relaxation times at 294 K, together with the data at ambient pressure [8]. The empirical formula $\tau_{CD} = \tau_0 \exp[DP/(P_0-P)]$ [12,17] provides a good fit of the $\alpha$-relaxation time (dashed line). It should be noted that the $\tau_{CC}$ results at $P < 3$ GPa have very high uncertainty as here no well-resolved $\beta$-peak is observed (cf. Fig. 1). Nevertheless, they well match the result at ambient pressure, obtained in [8] by modeling the EW with a CC function. This finding indicates that the secondary relaxation peaks observed at very high pressures are due to the same process as the relaxation causing the EW at 0.1 MPa. At $P > 3$ GPa, the behavior of $\tau_{CC}(T)$ dramatically changes and the $\alpha$ and $\beta$ time scales separate, which leads to the clear emergence of a separate $\beta$-peak in Figs. 1 and 2.

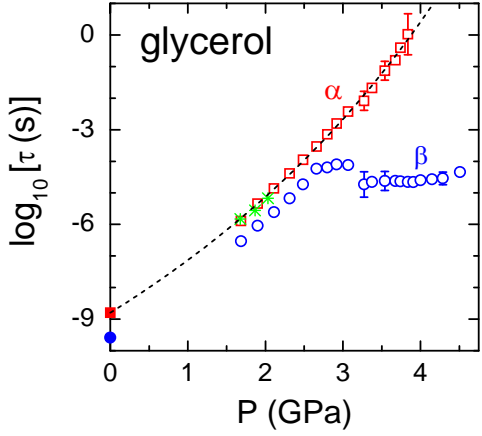

FIG. 3 (color online). Pressure dependences of the relaxation times for $\alpha$- (squares) and $\beta$-processes (circles) at 294 K. The stars show the $\alpha$-relaxation times published in [17]. The ambient-pressure results (closed symbols) were taken from [8]. The dashed line is a fit with $\tau = \tau_0 \exp[DP/(P_0-P)]$ ($\tau_0 = 1.6\times10^{-9}$ s, $D = 41.7$, and $P_0 = 11.9$ GPa) [12,17]. For $P < 3$ GPa, the increasingly large errors of the $\beta$-relaxation times are difficult to quantify.

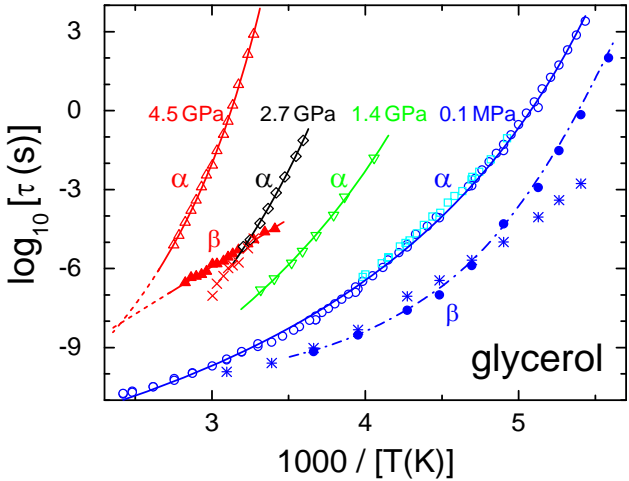

FIG. 4 (color online). Temperature dependences of relaxation times at four different pressures. The open symbols denote the $\alpha$-relaxation times $\tau_{CD}$. For 0.1 MPa, additional data from Ref. [3] are included (open circles). For 0.1 MPa [8] and 4.5 GPa, $\beta$-relaxation times $\tau_{CC}$ are provided (closed circles and triangles, respectively). The predictions for $\tau_\beta(T)$ from eq. (5) are shown as stars and crosses. The solid lines are fits using eq. (3), for $\tau_{CD}(T)$ and the Arrhenius law for $\tau_{CC}(T)$ at 4.5 GPa. The dash-dotted line is drawn to guide the eyes.

The temperature dependences of $\tau_{CD}$ and $\tau_{CC}$ at constant pressures are plotted in Fig. 4. At all pressures $\tau_{CD}(T,P=\text{const})$ is well described by the empirical modified Vogel-Fulcher-Tammann-Hesse formula [1,24]:

$$\tau_{CD} = \tau_0 \exp[DT_0/(T-T_0)]. \qquad (3)$$

By fitting $\tau_{CD}(T)$ to (3), the isobaric fragility (or steepness index) $m_P$ can be calculated [25] via

$$m_P = \left.\frac{d\log(\tau_{CD})}{d(T_g/T)}\right|_{T=T_g} = \frac{DT_0 T_g}{\ln(10)(T_g-T_0)^2}, \qquad (4)$$

where $D$ and $T_0$ are fitting parameters in (3), while $T_g$ is the temperature at which $\tau_{CD}(T) = 10^3$ s (Fig. 4). Fragility, which correlates with various dynamic properties, provides a useful classification of glass formers [12,19,25]. The pressure dependences of $m_P$ and $T_0$ and the glass transition temperature $T_g$, calculated by using data from Fig. 4, are presented in Fig. 5, together with data from Refs. [16,19,26,27]. Within experimental error, $T_0(P)$ and $T_g(P)$ are in good agreement with literature data. Concerning $m_P(P)$, our results signal a rather continuous increase, in contrast to an increase followed by a saturation beyond 1 GPa, reported in [16,27].

To investigate the origin of the secondary relaxation in glycerol, we consider $\tau_{CC}(T)$ in Fig. 4 for 0.1 MPa and 4.5 GPa. In both cases, at high temperature (e.g., for $\tau_{CD} \approx 10^{-5}$ s) the separation of $\alpha$- and $\beta$-relaxation dynamics is of similar magnitude. However, when approaching $T_g$ (at $\tau_{CD} \approx 10^3$ s), both relaxation times become separated by many decades for 4.5 GPa, which leads to a well-pronounced $\beta$-peak in Figs. 1 and 2, while they approximately retain the same separation at ambient pressure, where only an EW is observed. To understand this behavior, we consider a correlation of $\alpha$- and JG-relaxation times, which is based on empirical findings in an number of glass formers [28] and which is consistent with the coupling model [29], predicting

$$\tau_{JG} \equiv \tau_\beta \approx (t_c)^n (\tau_{KWW})^{1-n}. \qquad (5)$$

Here $\tau_{KWW}$ and $n = 1 - \beta_{KWW}$ are parameters of the Kohlrausch-Williams-Watts (KWW) function and $t_c \approx 2$ ps. This relation, which was found to hold for a variety of different glass formers [8,28], was proposed as a way to distinguish genuine JG relaxations from other types of secondary relaxations [6]. As shown in [8], within this framework the EW relaxation at ambient pressure is found to be of JG type. Its small splitting from the $\alpha$ time scale is consistent with $\tau_{JG}(T)$ from eq. (5) (stars in Fig. 4) and can be ascribed to the rather small values of $n$ in glycerol [8,30]. For 4.5 GPa we have calculated the KWW parameters from the corresponding CD function parameters [31]. $\beta_{CD}$ determined from the fits in Fig. 2 has very large uncertainty due to the restricted frequency range in these high-pressure experiments and the overlap of $\alpha$- and $\beta$-relaxation, especially at high temperatures. Thus, the results for $\tau_{JG}$ in Fig. 4 (crosses) can only provide a rough estimate and no significant data at $1000/T < 3$ K$^{-1}$ could be obtained. Nevertheless, the found agreement with the experimentally determined $\tau_{CC}$ at lower temperatures certainly is indicative of a JG-type relaxation. It should be noted that the pressure dependence of $\tau_{CC}$ (Fig. 3) is weak in comparison with those in van-der-Waals glass formers [6] because glycerol is hydrogen-bonded and behaves differently.

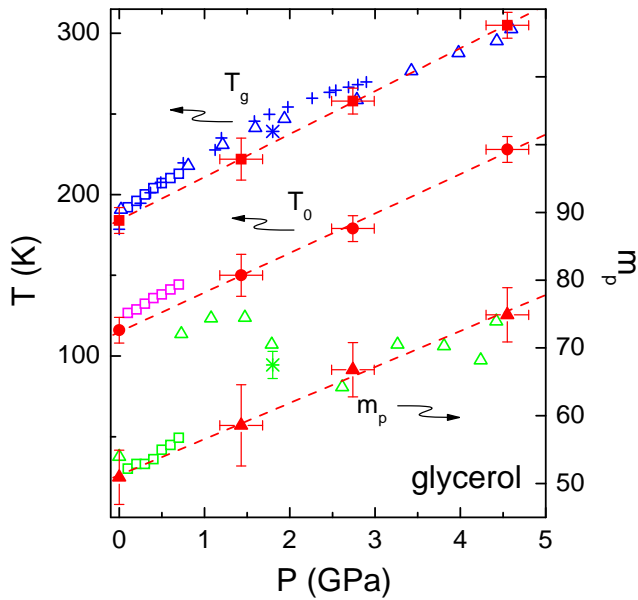

FIG. 5 (color online). Pressure dependences of $T_0$ (eq. 3) and $T_g$ (left axis) and of the isobaric fragility $m_p$ (eq. 4, right axis). The open squares at 0.1-0.7 GPa were taken from [19]. The crosses denote $T_g$ from viscosity measurements [26]. The open triangles show $T_g$ and $m_p$ from [16]. The single points at 1.8 GPa (stars) are $T_g$ and $m_p$ from [27].

The final question is, why $\alpha$- and $\beta$-relaxation separate so much more at high pressures (Fig. 4). This may be explained by a breakdown of the hydrogen bonds under high pressure. Then one may speculate about a stronger coupling of the molecules, which within the coupling model [28] should lead to an increase of the coupling parameter $n$, i.e. a decrease of $\beta_{KWW}$. Via eq. (5), this leads to a stronger separation of $\tau_{JG}$ and the $\alpha$-relaxation time. This scenario is also consistent with the increase of the fragility (Fig. 5), which should be larger for hard-sphere like glass formers and, in addition, generally is larger for smaller values of $\beta_{KWW}$ [25], i.e. for larger $n$. However, admittedly these statements are somewhat speculative as the high uncertainty of $\beta_{CD}$ or $\beta_{KWW}$, mentioned above, prevents an unequivocal determination of the actual pressure dependence of $n$.

In summary, we have applied a method for obtaining high-quality spectra at superhigh pressures to the study of the archetypical glass-forming liquid glycerol at pressures up to 4.5 GPa. A distinct secondary relaxation, most likely being of JG-type, has been discovered at pressures above 3 GPa. The pressure-induced enhancement of the coupling between molecules most likely induces the separation of the EW relaxation from the $\alpha$-peak. The present results demonstrate that superhigh pressure experiments can provide access to new exiting phenomena in glassy matter.

The authors wish to thank G.P. Johari, K.L. Ngai, V.N. Ryzhov, S.M. Stishov, and O.B. Tsiok for valuable discussions. This work has been supported by the RFBR Russian-German Joint Program (09-02-91351), by the Programs of the Presidium of RAS, and by the Deutsche Forschungsgemeinschaft via the German-Russian joint research project, Grant-No. 436 RUS113/992/0.

___________________

*Corresponding author. Email address: Peter.Lunkenheimer@Physik.Uni-Augsburg.de